\documentclass[pra,aps,eqsecnum,amsmath]{revtex4}
\begin{document}

\title{Non-quantum Entanglement and a Complete Characterization of 
pre-Mueller and Mueller Matrices in Polarization Optics}
\author{B. Neethi Simon}
\affiliation{Department of Applied Mechanics, IIT Madras, 
Chennai 600~036}
\author{Sudhavathani Simon}
\affiliation{Department of Computer Science, Women's Christian College, 
Chennai 600~006}
\author{N. Mukunda}
\affiliation{Centre for High Energy Physics, Indian Institute of 
Science, Bangalore 560~012}
\author{F. Gori and Massimo Santarsiero}
\affiliation{Dipartimento di Fisica, Universit\'{a} Roma Tre and 
CNISM, Vila della Vasca Navale~84, I-00146 Rome}
\author{Riccardo Borghi}
\affiliation{Dipartimento di Electronica Applicata, Universit\'{a} Roma 
Tre and 
CNISM, Vila della Vasca Navale~84, I-00146 Rome}
\author{R. Simon}
\affiliation{The Institute of Mathematical Science, Tharamani, Chennai 
600~113}


\begin{abstract}
The Mueller-Stokes formalism which governs conventional polarization 
optics is formulated for plane waves, and thus the only 
qualification  one could 
 demand of a $4\times 4$ real matrix $M$ in order that it 
 qualifies to be the Mueller matrix of some physical system is that 
$M$ should map 
$\Omega^{({\rm pol})}$, 
the positive cone of Stokes vectors, into 
itself. In view of growing current interest in the characterization of 
partially 
coherent partially polarized electromagnetic beams, there is  need to 
extend this formalism to such beams wherein the polarization and spatial 
dependence are generically inseparably intertwined. This inseparability 
or non-quantum entanglement brings in additional constraints that a 
pre-Mueller matrix $M$  mapping  $\Omega^{({\rm pol})}$ into itself 
needs to meet in order that it is an acceptable physical  Mueller 
matrix. 
These additional constraints are motivated and fully characterized.  

\end{abstract}
\maketitle

%
\baselineskip14pt
\section{Introduction}

Entanglement is traditionally studied almost exclusively in the 
context of quantum 
systems. However,  this notion is basically kinematic in nature,  
 and so  is bound to present itself  whenever and wherever  
the state space of interest is the  tensor product of two (or more) 
vector spaces. The  vectors of the individual spaces, and hence 
(tensor) 
products of such vectors, will be expected to possess identifiable 
physical meaning. 
 Polarization optics of paraxial electromagnetic beams happens to 
have  
precisely 
this kind of a setting, and so one should expect entanglement 
 to play a  nontrivial role in this situation. It turns out that  
 entanglement in this  non-quantum 
setup is not just a matter of academic curiosity: we shall show in 
this 
 paper that {\em this non-quantum entanglement 
 helps in resolving a fundamental issue in classical 
polarization optics}. It will appear that this issue could not have been 
resolved without explicit consideration of entanglement.

A paraxial beam propagating along the positive $z$-axis is completely 
 determined  in terms 
of the transverse components of the electric field specified  throughout 
a  transverse plane $z=$ constant as functions of  the transverse 
variables $(x,\,y)= \mbox{\boldmath$\rho$}$. 
 If these components are independent of the transverse coordinates, then  
the situation corresponds to a plane wave propagating along the $z$-axis. 
The traditional Mueller-Stokes formalism in terms of Stokes vector $S$ 
and Mueller matrix $M$, describing respectively the beam and the optical 
system, presumes essentially this kind of situation wherein the spatial 
degree of 
freedom can be safely left out of consideration, the focus being on the 
polarization degree of freedom\,\cite{MandelWolf,Brosseau,Gil07}.  

Recent years have witnessed an enormous interest in partially polarized 
partially coherent electromagnetic 
beams\,\cite{James94,Gori98,Gorikids98,Gori01,Gori03,Wolf03,Tervo03,Setala04,Dennis04,Refregier05,Korotkova05,Gori07,Luis07,Mejias07a,Mejias07b} 
and hence there is a need to 
extend the Mueller-Stokes formalism to such beams. 
 Given a $4\times 4$ real matrix $M$,  
 it should map $\Omega^{({\rm pol})}$, 
the positive cone of Stokes vectors, into 
itself in order that it could be the Mueller matrix of some physical 
system. {\em Within the conventional formalism, this seems to be the only 
qualification that can be demanded of $M$}. In a partially coherent 
partially  polarized beam, polarization and spatial 
dependence happen to be  inseparably intertwined. 
This inseparability 
or non-quantum entanglement brings in additional constraints that a 
pre-Mueller matrix $M$  mapping  $\Omega^{({\rm pol})}$ into itself 
needs to meet in order that it is a physically acceptable Mueller 
matrix. 

 The principal purpose of the present work is to motivate these 
constraints 
and characterize them fully.   
 The next two Sections of the paper act as preparation towards this end. 
 We begin  in Section~2 by recounting the Mueller-Stokes formalism  
 as it applies to plane waves.  This is  then extended in Section~3  
 to paraxial electromagnetic beams, and 
 the role of (non-quantum) entanglement between polarization and spatial 
modulation is rendered transparent. 
These two Sections equip us with all the tools needed to formulate in 
Section~4 the additional physical constraints on a pre-Mueller 
matrix $M$,  
arising as consequence of entanglement. Our final result is 
formulated in the form of a necessary and sufficient condition, and 
a simple illustrative example is treated in some detail for illustration 
of the nature 
of 
these further constraints. And we conclude in Section V with some further 
remarks. 

\section{Polarization optics of plane waves}
 
For a plane wave whose propagation direction is along the (positive) 
$z$-axis perpendicular to the $(x,\,y)$-plane, the  
$x$ and $y$
components $E_1,\,E_2$ of the electric field are independent of 
the transverse-plane coordinates   $\mbox{\boldmath $\rho$}$, 
and  can be arranged into a (numerical) {\em column vector}
\begin{eqnarray}
 {\mbox{\boldmath$E$}}
\equiv  
\left[\begin{array}{c} E_1\\ E_2\end{array}\right]\in
{\cal C} ^2.
\end{eqnarray}
We have suppressed, for convenience, a space-time dependent scalar 
factor of the form
$e^{i(kz-\omega t)}$. 
 While ${\mbox{\boldmath$E$}}^{\dagger}{\mbox{\boldmath$E$}}
 =|E_1|^2+|E_2|^2$ is (a
measure of) the intensity, the ratio $\gamma=E_1/E_2$ of the (complex)
components, which ratio can be viewed as a point on the Riemann or 
Poincar\'{e} sphere 
$S^2$,
specifies the polarization state. In particular, the signature of the
imaginary part of $\gamma $ describes the handedness of the 
(generally elliptic) polarization.

In presence of fluctuations ${\mbox{\boldmath$E$}}$  acquires some 
randomness, and in
this case the state of polarization  is effectively described by the
 $2\times 2$ {\em coherency or polarization matrix}
\begin{eqnarray}
 \Phi  \equiv \langle 
{\mbox{\boldmath$E$}}
{\mbox{\boldmath$E$}}^{\dagger}\rangle 
=\left[\begin{array}{cc}
\langle E_1E_1^*\rangle &
\langle E_1E_2^*\rangle\\
\langle E_2E_1^*\rangle&
\langle E_2E_2^*\rangle
\end{array}\right],
\end{eqnarray}
where $ \langle\,\cdots\,\rangle$ denotes ensemble average. The 
coherency 
matrix
is hermitian, $\Phi^{\dagger}=\Phi$, 
and positive semidefinite,
${\mbox{\boldmath$V$}}^{\dagger}\Phi {\mbox{\boldmath$V$}}
={\rm tr}\,(\,\Phi\, 
{\mbox{\boldmath$V$}}{\mbox{\boldmath$V$}}^{\dagger}\,)
\ge 0, \;\,\forall\, {\mbox{\boldmath$V$}}\in
{\cal C} ^2$. This positivity property may be denoted simply  
$\Phi\ge0$.  Hermiticity and positivity 
  are the {\em defining properties} of $\Phi$\,: every $2\times2$ 
matrix obeying these two conditions is a valid coherency matrix, and 
represents some  polarization state.  Since $\Phi$ is a $2\times 2$ 
matrix, the positivity condition takes the simple scalar form 
\begin{eqnarray}
{\rm tr}\,\Phi &>&0\,,\nonumber\\
\det\,\Phi&\ge& 0\,.
\end{eqnarray}
It is clear that the intensity corresponds to ${\rm tr}\,\Phi$. 
Fully polarized light (pure states) corresponds to ${\rm det}\,\Phi=0$, 
and  partially
 polarized or mixed states 
correspond to $\det\,\Phi>0\,.$ 

{\em Typical systems of interest in polarization optics are 
  transversely homogeneous}, in the sense that their action is 
independent 
of the coordinates $(x,\,y)$ spanning the transverse plane in which the 
 system  lies. If such a system is 
deterministic and acts
linearly at the field amplitude level, it is described by a complex
$2\times 2$ numerical matrix $J$ called the Jones matrix of the system:
\begin{eqnarray}
 J&=&\left[\begin{array}{cc}
J_{11}& J_{12}\\
J_{21}& J_{12}\end{array}\right]
:\;\;\;
{\mbox{\boldmath$E$}}
   \to {\mbox{\boldmath$E$}}'
 = \left[\begin{array}{c} E'_1\\ E'_2\end{array}\right] 
=J{\mbox{\boldmath$E$}}\nonumber\\
    &&\Leftrightarrow ~~ \Phi
  \equiv 
\langle {\mbox{\boldmath$E$}}
{\mbox{\boldmath$E$}}^{\dagger}\rangle 
\to \Phi' = \langle {\mbox{\boldmath$E$}}^{\,\prime}
{\mbox{\boldmath$E$}}^{\,\prime\,\dagger}\rangle 
=J\Phi J^{\dagger}.~~\;
\end{eqnarray}
It is clear that Jones systems map pure states ($\det\,\Phi = 0$) into 
pure states.

Since $\Phi$ is hermitian, it can be conveniently described as 
{\em real} linear combination of the four orthogonal hermitian 
matrices 
$\tau_0
= 1_{2\times 
2},\;\tau_1=\sigma_3,\;\tau_2=\sigma_1,\;\tau_3=\sigma_2$\,: 
\begin{eqnarray}
 \Phi=\frac{1}{2}\sum_{a =0}^{3}S_{a}\tau_{a} ~\Leftrightarrow ~
S_{a}={\rm tr}(\tau_{a}\Phi); ~ {\rm tr}\,\tau_a\tau_b = 
2\delta_{ab}.\;\;
\end{eqnarray}
The reason for 
choosing the $\tau$-matrices, a permuted version of the Pauli matrices  
rather than the Pauli $\sigma$-matrices themselves, is to be consistent 
with the  optical 
convention that the circularly polarised states, the eigenstates of 
$\sigma_2$, be along the `third' 
axis (polar axis) of the Poincar\'{e} sphere. 
The intensity equals $S_0={\rm tr}\,\Phi$. The
expansion coefficients $S_{a} $ are the components of the {\em Stokes 
vector} $S\in \mbox{\boldmath$R$} ^4$. Note that $\tau_3^*= -\tau_3$ 
and $\tau_a^* =\tau_a$ if $a\ne 
3$\,.   

While hermiticity of $\Phi$  is equivalent to reality of
$S \in \mbox{\boldmath$R$}^4$, the positivity conditions ${\rm 
tr}\,\Phi>0,~{\rm det}\,\Phi\ge0$
read, respectively, 
\begin{eqnarray}
 S_0&>&0,\nonumber\\
S_0^2-S_1^2-S_2^2-S_3^2&\ge& 0.
\end{eqnarray}
Thus, {\em permissible polarization states correspond to the 
positive light
cone and its interior} (solid cone). Pure states live on the surface 
of
this cone. As suggested by this  light cone structure, the {\em proper 
orthochronous 
Lorentz
group} $SO(3,1)$ plays  quite an important role  in polarization 
optics\,\cite{Barakat81,Sridhar94}.

Under the action of a deterministic or Jones system $J$ 
 described in (2.4) the elements of the output coherency matrix 
$\Phi'$ are obviously linear in those of $\Phi$. This, in view of the 
linear
relation (2.5) between $\Phi$ and $S$, implies that under passage 
through
such a system the output Stokes vector $S'$ and the input $S$ will be
linearly related by a $4\times 4$ real matrix $M(J)$ determined by
$J$:
\begin{eqnarray}
 J:\;S \to S'=M(J)S.
\end{eqnarray}
We may call $M(J)$ the Mueller matrix of the Jones system $J$. It is
known also as a Mueller-Jones matrix to emphasise the fact that it 
is constructed out of a Jones matrix. While $\Phi =
\langle {\mbox{\boldmath$E$}}
{\mbox{\boldmath$E$}}^{\dagger}\rangle$ is a $2\times 2$  matrix, 
the tensor product  $\tilde{\Phi} 
\equiv 
\langle {\mbox{\boldmath$E$}}
\otimes {\mbox{\boldmath$E$}}^*\rangle$ is a four-dimensional column vector 
associated with $\Phi$: 
\begin{eqnarray}
\tilde{\Phi}= \left[\begin{array}{c} \tilde{\Phi}_0\\
    \tilde{\Phi}_1\\ \tilde{\Phi}_2\\
    \tilde{\Phi}_3\end{array}\right]
\equiv 
 \left[\begin{array}{c} 
\langle E_1E_1^*\rangle\\
\langle E_1E_2^*\rangle\\
\langle E_2E_1^*\rangle\\
\langle E_2E_2^*\rangle\\
\end{array}\right]
=
\left[\begin{array}{c} {\Phi}_{11}\\
    {\Phi}_{12}\\ {\Phi}_{21}\\
    {\Phi}_{22}\end{array}\right].
\end{eqnarray}
This idea of going from a pair of indices, each running over $1$ and $2$, 
to a single index running over $0$ to $3$ and vice versa can often be 
used  to advantage to associate with any $2\times 2$ matrix $K$ a 
corresponding column vector 
$\tilde{K}$ with  
$\tilde{K}_0 = K_{11}$, $\tilde{K}_1 = K_{12}$,
 $\tilde{K}_2 = K_{21}$, and 
$\tilde{K}_3 = K_{22}$.  
The tensor product $J\otimes J^*$ is a $4\times4$ matrix:
\begin{eqnarray}
J\otimes J^* &\equiv&  
\left[\begin{array}{cc} 
J_{11}J^*& J_{12}J^*\\
J_{21}J^*& J_{22}J^*\end{array}\right]\nonumber\\
&=&
\left[\begin{array}{cccc} 
J_{11}J_{11}^*& J_{11}J_{12}^*&J_{12}J_{11}^*& J_{12}\,J_{12}^*\\
J_{11}J_{21}^*& J_{11}J_{22}^*&J_{12}J_{21}^*& J_{12}\,J_{22}^*\\
J_{21}J_{11}^*& J_{21}J_{12}^*&J_{22}J_{11}^*& J_{22}\,J_{12}^*\\
J_{21}J_{21}^*& J_{21}J_{22}^*&J_{22}J_{21}^*& J_{22}\,J_{22}^*
\end{array}\right].~~\;
\end{eqnarray}
The transformation $\Phi\to \Phi'=J\Phi J^{\dagger} $ is thus 
equivalent to $\tilde{\Phi}\to\tilde{\Phi}'=J\otimes J^*\tilde{\Phi}.$
 Since  $\tilde{\Phi}$ is related to the Stokes vector through 
\begin{eqnarray}
\left[\begin{array}{c}S_0\\S_1\\S_2\\S_3\end{array}\right]=
\left[\begin{array}{cccc} 1&0&0&1\\ 1&0&0&-1\\ 0&1&1&0\\
    0&i&-i&0\end{array}\right] 
\left[\begin{array}{c} \tilde{\Phi}_0\\
    \tilde{\Phi}_1\\ \tilde{\Phi}_2\\
    \tilde{\Phi}_3\end{array}\right],
\end{eqnarray}
it follows that 
\begin{eqnarray}
M(J)=A(J\otimes J^*)A^{-1},
\end{eqnarray}
$A$ being the  $4\times 4$ matrix exhibited in Eq.\,(2.10); this
matrix is {\em essentially} unitary:
$A^{-1}=\frac{1}{2}A^{\dagger}$.

If ${\rm det}\,J$ is of unit magnitude, then $M(J)$  computed by this
prescription is an element of $SO(3,1)$, the proper orthochronous group of 
Lorentz transformations;  this was to be expected 
in view of the {\em two-to-one homomorphism} between $SL(2,C)$ and
$SO(3,1)$. It follows that for any nonsingular $J$  
 the associated Mueller-Jones matrix $M(J)$ is $|{\rm det}\,J|$ times 
an 
element of $SO(3,1)$. {\em The prescription (2.11), though, applies to 
singular Jones matrices as well}.

\subsection{Mueller matrices arising from Jones systems}

A nondeterministic system is described {\em directly} by a Mueller matrix
$M:\;S\to S'=MS$ and,  by definition, such a Mueller matrix cannot  equal
$M(J)$ for any $2\times2$ (Jones) matrix $J$.  
Given a Mueller matrix $M$, how to test if it is a Mueller-Jones
matrix  for some $J$ or, equivalently, how to test if the system
described by $M$ is a deterministic or Jones system? It turns out that 
this question which has received much 
attention\,\cite{Stokes,Abhyankar69,Fry81,Barakat81} 
has a simple and elegant solution {\cite{Simon82}}. 

We go over in some detail the construction underlying this solution, 
for 
it plays a central role in our analysis to follow. 
The sixteen
$4\times 4$ {\em hermitian} matrices 
$ U_{ab} = 
\frac{1}{2}\,\tau_{a}\otimes \tau_{b}^{*}$, with 
$a,\,b$
independently running over the index set $\{0,\,1,\,2,\,3\}$, form an 
orthonormal 
set or basis in the (vector) space of $4\times 4$ matrices; these matrices 
are {\em unitary} and {\em self-inverses}\,: 
\begin{eqnarray}
 U_{ab} &=& 
\frac{1}{2}\,\tau_{a}\otimes \tau_{b}^{*} =U_{ab}^{\dagger} 
=U_{ab}^{\,-1}\,,\nonumber\\ 
{\rm tr}\,(\,U_{ab}U_{cd}\,) &=&\delta_{ac}\delta_{bd}\,,
~ a,b,c,d \in \{1,2,3,4\}.~~\;
\end{eqnarray}
[\,Complex conjugation of the second factor of the tensor product 
$\tau_a\otimes \tau_b^*$ is suggested by the construct $J\otimes J^*$ in 
   (2.11)\,]. 
 Thus every 
({\em hermitian}) $4\times 4$ matrix can be written 
uniquely as a  ({\em real}) linear combination of
$\{\tau_{a}\otimes\tau_{b}^{*}\}$.  
Aa an important consequence of this fact we have\,:

\noindent
{\bf Proposition}~1\,: There exists  a natural one-to-one 
correspondence between the set of all $4\times 4$ real matrices 
and the set of all $4\times 4$ hermitian matrices.

 Indeed, a real matrix  $M$ and the associated  hermitian matrix 
$H_M$ are related in this simple manner\,: 
\begin{eqnarray}
H_M=\frac{1}{2}\sum_{a,\,b=0}^3
M_{ab}\tau_{a}\otimes\tau_{b}^*\,. 
\end{eqnarray}
Similarly, a hermitian matrix $H$ and the associated real matrix $M_H$ 
are related through
\begin{eqnarray}
(\,M_H\,)_{ab}=\frac{1}{2}{\rm
  tr}\,(H\tau_{a}\otimes\tau_{b}^*)\,, ~~~a,\,b= 0,\,1,\,2,\,3\,.
\end{eqnarray}
It is clear that these relations are inverses of one another. 
We write these in more detail for  later 
use: 
\begin{eqnarray}
H_M=\frac{1}{2}
\left[
\begin{array}{llll}
M_{00} + M_{11} & 
M_{02} + M_{12} & 
M_{20} + M_{21} & 
M_{22} + M_{33} \\ 
~~+  M_{01} +  M_{10} & 
 ~+ i(M_{03} +  M_{13})~~& 
 ~- i(M_{30} +  M_{31})~~\;& 
 ~+ i(M_{23} -  M_{32})~~\\ 
 & & & \\
M_{02} + M_{12} & 
M_{00} - M_{11} & 
M_{22} - M_{33} & 
M_{20} - M_{21} \\
~ - i(M_{03} +  M_{13})~~& 
~~ -   M_{01} +  M_{10} & 
~ - i(M_{23} +  M_{32})~~& 
~ - i(M_{30} -  M_{31})~~\\
& & & \\
M_{20} + M_{21} & 
M_{22} - M_{33} & 
M_{00} - M_{11}  & 
M_{02} - M_{12} \\ 
~+ i(M_{30} +  M_{31})~~& 
~ + i(M_{23} +  M_{32})~~& 
~~ +   M_{01} -  M_{10} & 
~ + i(M_{03} -  M_{13})~~\\ 
   & & & \\
M_{22} + M_{33} & 
M_{20} - M_{21} &
M_{02} - M_{12} &
M_{00} + M_{11} \\  
~ - i(M_{23} -  M_{32})& 
~ + i(M_{30} -  M_{31})&
~ - i(M_{03} -  M_{13})&
~~ -   M_{01} -  M_{10}  
\end{array}
\right].\nonumber\\
\end{eqnarray}
This matrix in identical form was first presented in\,\cite{Simon82}.  
Of the sixteen $4\times 4$  matrices $U_{ab}$, only 
$U_{20},\, U_{21},\, U_{30}$  and $U_{31}$ have  
{\em nonzero entries} at the $_{1 3}$ location, and this 
explains the entry 
$M_{20} -M_{21} -i(M_{30} -M_{31})$ 
 for $(\,H_M\,)_{13}$. Written in detail, the 
relation (2.14) has the form   
\begin{eqnarray}
M_H=\frac{1}{2}
\left[
\begin{array}{llll}
H_{00} + H_{11} & 
H_{00} - H_{11} & 
H_{01} + H_{10} & 
-i(H_{01} - H_{10})\\
~~ +   H_{22} +  H_{33} & 
 ~~+   H_{22} -  H_{33} & 
 ~~+   H_{23} +  H_{32} & 
~-i(H_{23} -  H_{32})\\
 & & & \\
H_{00} + H_{11} & 
H_{00} - H_{11} & 
H_{01} + H_{10} & 
-i(H_{01} - H_{10})\\
~~-   H_{22} -  H_{33} & 
~~-   H_{22} +  H_{33} & 
~~-   H_{23} -  H_{32} & 
+i(H_{23} -  H_{32})\\
  & & & \\ 
H_{02} + H_{20} & 
H_{02} + H_{20} & 
H_{03} + H_{30} & 
 -i(H_{03} - H_{30})\\
~~+   H_{13} +  H_{31} & 
~~-   H_{13} -  H_{31} & 
~~+   H_{12} +  H_{21} & 
~ +i(H_{12} - H_{21})\\
 &  & &  \\
 i(H_{02} - H_{20}) & 
 i(H_{02} - H_{20}) & 
  i(H_{03} - H_{30})&
 H_{03} + H_{30}  \\
 ~+i(H_{13}  - H_{31})~~ & 
 ~-i(H_{13}  - H_{31})~~ & 
 ~+i(H_{12} -  H_{21})~~&
 ~~-   H_{12} -  H_{21}
\end{array}
\right]. \nonumber\\
\end{eqnarray}
The entry $ -i(H_{01} -H_{10}) +i(H_{23} -H_{32})$  
for $(\,M_H\,)_{13}$ is explained by the fact that 
the nonzero entries of $U_{13}$ are at the $_{01},\,_{10},\,_{23}$ and 
$_{32}$ 
locations. 

Fundamental to the structure of the Mueller-Stokes formalism 
of polarization optics is the following result\,:

\noindent
{\bf Proposition} 2\,: 
Given a $4\times 4$ real matrix $M$, it is a Mueller-Jones matrix iff 
the associated  hermitian matrix $H_M$ is a one-dimensional
projection. That is iff  $H_M =\tilde{J}\tilde{J}^{\dagger}$ 
 for some (complex) four-dimensional column vector $\tilde{J}$. If 
$H_M =\tilde{J}\tilde{J}^{\dagger}$, then  the $2\times2$ matrix $J$ 
 associated with $\tilde{J}$ is 
the Jones matrix of the deterministic system represented by $M$.

In the original work\,\cite{Simon82} where 
 this proposition was formulated and proved,  this hermitian matrix
$H_M$ was actually denoted $N$. However, in the present work we use 
 instead the symbol $H$ to emphasise 
 hermiticity, the most important property of this matrix.

Consider now a  transformation which is a {\em convex sum} of 
Jones systems:
\begin{eqnarray}
 \Phi \to \Phi' =\sum_{k=1}^np_kJ^{(k)}\Phi {J^{(k)}}^{\dagger},
~p_k>0,~\sum_{k=1}^{n} p_k=1.~~\;
\end{eqnarray}
This transformation may be realized by a set of $n$ deterministic or 
Jones systems $J^{(1)},\,J^{(2)},\,\cdots,\,J^{(n)}$  
arranged in parallel,  with a fraction $p_k$ of the light going
through the $k^{{\rm th}}$ Jones system $J^{(k)}$, and all the
transformed beams combined (incoherently) at the output. 
It can also be viewed as a fluctuating  system which assumes the 
Jones form 
$J^{(k)}$ with probability $p_k$. In either case, 
 it is clear that  the Mueller
matrix $M$ of this nondeterministic system and its associated hermitian
matrix are {\em corresponding convex sums}:
\begin{eqnarray}
M=\sum_{k=1}^{n}p_kM(J^{k}),
~~H_M=\sum_{k=1} ^{n} p_k\tilde{J}^{(k)}\tilde{J}^{(k)\,\dagger}.
\end{eqnarray}

It is useful to denote by $\{\,p_k,\,J^{(k)}\,\}$ the convex sum 
 or ensemble realization 
represented by Eq.\,(2.17), or equivalently by Eq.\,(2.18).  
  Obviously, such an ensemble or convex sum realization 
 always leads to a positive semidefinite $H_M$, and  
 positive semidefinite $H$ alone can be realized as a convex sum of 
projections. Thus  as an immediate, and mathematically trivial,  
consequence of 
Proposition~2 we have

\noindent
{\bf Corollary}\,: An optical system described by $M$ is 
realizable as a convex sum or ensemble $\{\,p_k,\,J^{(k)}\,\}$ of Jones 
systems 
iff the associated $H_M\ge 0$.  If $H_M\ge 0$ the  number 
of Jones systems, $n$,  needed  for such a realization satisfies 
$n\ge r$, where $r$ is the rank of $H_M$. There is no upper limit on $n$ 
if $r\ge 2$.  

This corollary is physically important, and has attracted considerable 
attention\,\cite{Kim-Mandel-Wolf,Gil85,Gil87,Cloude89,Mee93,Anderson94,Woerdman07,Sudha08}.

 \subsection{Pre-Mueller matrices and their classification}

Given a  $4\times 4$ real matrix $M$, 
 Proposition~2 gives the necessary  and 
sufficient condition for $M$ to arise as 
the Mueller matrix of some Jones system $J$. 
  That still leaves open this 
 more general question: how to ascertain if a given 
matrix $M$ is a Mueller matrix? This question has an interesting 
history which is surprisingly recent. 

In traditional polarization optics, which is formulated for plane waves 
and not for beams,  
the state space  $\Omega^{({\rm pol})}$ is the collection of all Stokes 
vectors:
\begin{eqnarray} 
\Omega^{({\rm pol})} &=& \{\,S\in 
\mbox{\boldmath$R$}^{\,4}\,\,
|\,\,S_0>0,\;\;S^TGS\ge0\,\}\,,
\nonumber\\
G&=&{\rm diag}\,(1,\,-1,\,-1,\,-1)\,\nonumber\\
S^TGS&=&S_0^2-S_1^2-S_2^2-S_3^2\,.
\end{eqnarray} 
Now $G$ is the `Lorentz metric', and thus the state space 
$\Omega^{({\rm pol})}$ 
is the positive ({\em solid}) light 
cone in $\mbox{\boldmath$R$}^{\,4}$. Since a physical Mueller matrix 
should necessarily  map 
states 
into states, our 
question reduces to one of effectively characterising real linear 
transformations in $\mbox{\boldmath$R$}^{\,4}$ which map the positive 
 (solid) light cone {\em into} itself. 
 While $SO(3,1)\,\cup\,GSO(3,1)$, where  $SO(3,1)$ is  the 
proper 
orthochronous Lorentz group $SO(3,1)$, is 
the 
answer in the case of {\em onto} maps, the {\em more general into case}  
 was raised in Ref.\,\cite{Sanjay92} as a serious issue in 
polarization optics. This issue was formulated as two simple conditions 
that  $M$ has to meet  
[\,Eqs.\,(2.29),\,\,(2.31) of Ref.\,\cite{Sanjay92}\,],
 corresponding to  
the demand that the intensity and degree of polarization of the output 
be physical {\em for every input pure state}. Further,    
 the measured Mueller matrices of 
Howell\,\cite{Howell79} were tested for  these conditions and 
violation was found in excess of 10\%\,, a magnitude considerably 
larger  than the tolerance 
suggested by the reported measurements. 
 Ref.\,\cite{Sanjay92} thus concluded that {\em the Howell system  
 fails to map the positive light cone $\Omega^{({\rm pol})}$ into 
itself}, and  
 this is {\em possibly  the first time} that a verdict of this kind    
 was made on some published Mueller matrices.  

 Subsequent progress in respect of this issue was quite rapid.  
 In a significant step forward  
  Givens and Kostinski\,\cite{Givens93} derived, based on an 
impressive analysis of the spectrum of $GM^TGM$, what appeared to be a 
necessary and 
sufficient condition for $M$ to map  
$\Omega^{({\rm pol})}$ 
into itself. They analysed the Howell system based on their 
own condition 
and concluded that their results were ``{\em in coincidence with the 
negative verdict on the Howell matrix delivered in} 
\cite{Sanjay92}''.   
Soon after,  
van der Mee\,\cite{Mee93} derived a more complete set of necessary 
and 
sufficient  conditions for $M$ to map $\Omega^{({\rm pol})}$ 
into itself;  the analysis of van der Mee too was based 
on the spectrum of $GM^T GM$.  

Decomposition of a  Mueller matrix $M$ in various product forms,  
to gain insight into the physical effects $M$ could have 
on the input polarization state, 
has been an activity of considerable 
interest\,\cite{Lu96,Ossikovski07,Devlaminck08,Boulvert09}.  
The importance of obtaining the canonical or normal forms 
of Mueller matrices under the double-coset transformation 
$M \to L_\ell ML_r,\;L_\ell,L_r\in SO(3,1)$ was motivated in 
Ref.\,\cite{Sridhar94}, and it was shown that the theorem of Givens 
and 
Kostinski  implied that  the canonical form of every 
nonsingular (real) matrix $M$ 
which maps $\Omega^{({\rm pol})}$ into itself is diagonal; i.e.,  
$M= L_\ell M^{(1)}L_r$ where $M^{(1)}= {\rm 
diag}\,(d_0,\,d_1,\,d_2,\,d_3),\;d_0 \ge d_1\ge d_2\ge|d_3|$ and  
 $L_\ell,L_r\in SO(3,1)$. 
 It turned out that while the result of van der Mee  is 
essentially complete\,\cite{Rao98a}, that 
of Givens and Kostinski is incomplete. This means the diagonal form 
$M^{(1)}$ of Ref.\,\cite{Sridhar94} noted  above is 
not the only canonical form  for a nonsingular $M$ mapping  
$\Omega^{({\rm pol})}$ 
into itself; there exists a non-diagonal canonical form $M^{(2)}$, and 
this is the case that was missed by the `theorem' of Givens and 
Kostinski quoted above.  In a remarkably impressive and detailed 
study Rao et 
al.\,\cite{Rao98a,Rao98b} have further explored and completed 
 the analysis of van 
der Mee, leading to a complete solution to the 
question of canonical form for Mueller matrices, under 
double-coseting by $SO(3,1)$ elements,  raised in 
Ref.\,\cite{Sridhar94}. 

Since these canonical forms play a key role 
in our analysis below, we list them here in a concise form. 
 Matrices $M$ which map 
the state space $\Omega^{({\rm pol})}$ into itself 
 divide into two major and two minor families\,:
\begin{eqnarray}
~{\rm Type~I}: && M = L_\ell M^{(1)}L_r, \;\;
L_\ell,\,L_r\in SO(3,1)\,,\nonumber\\
             && M^{(1)}= {\rm 
diag}\,(d_0,\,d_1,\,d_2,\,d_3),\,\nonumber\\
&&~~~~~~~d_0 \ge d_1\ge d_2\ge |d_3|\,;\nonumber\\  
{\rm Type~II}: && M = L_\ell M^{(2)}L_r, \;\;
L_\ell,\,L_r\in SO(3,1)\,,\nonumber\\
   && M^{(2)} =\left[\begin{array}{cccc}
d_0 & d_0-d_1 & 0 & 0\\
0 & d_1 & 0 & 0\\
0 & 0 & d_2 & 0\\
0 & 0 & 0 & d_3
\end{array}\right],\nonumber\\
&& d_0>d_1>0,\; \sqrt{d_0d_1} \ge d_2 \ge |d_3|\,;\nonumber\\
{\rm Polarizer}: && M = L_\ell M^{({\rm pol})}L_r, \;\;
L_\ell,\,L_r\in SO(3,1)\,,\nonumber\\
   && M^{({\rm pol})} = \left[\begin{array}{cccc}
d_0 & d_0 & 0 & 0\\
d_0 & d_0 & 0 & 0\\
0   & 0   & 0 & 0\\
0   & 0   & 0 & 0
\end{array}\right],\;d_0>0\,;\nonumber\\
{\rm Pin~Map}: && M = L_\ell M^{({\rm pin})}L_r, \;\;
L_\ell,\;L_r\in SO(3,1)\,,\nonumber\\
   && M^{({\rm pin})} =\left[\begin{array}{cccc}
d_0 & 0 & 0 & 0\\
d_0 & 0 & 0 & 0\\
0   & 0 & 0 & 0\\
0   & 0 & 0 & 0
\end{array}\right],\;d_0>0\,.\;
\end{eqnarray}
Since elements of $SO(3,1)$ have unit determinant, it follows that 
$d_3$ in  the 
Type-I and Type-II cases is positive, negative, or zero according as 
${\rm det}\,M$ is positive, negative, or zero. 
 The $M$ matrices in the third and fourth families are manifestly 
singular. 
The third family 
is a Jones system, the associated $H$ matrix being a projection; 
indeed, $M^{({\rm pol})}$ corresponds to 
a Jones matrix $J$ whose only nonvanishing element is $J_{11} 
=\sqrt{2d_0}$.  Finally, the PinMap family is named so 
because   $M^{({\rm pin})}$ produces a {\em fixed} output 
polarization state independent of the  input, the intensity 
of the output being independent of the state of polarization of the 
input.  This may be contrasted with $M^{({\rm pol)}}$: while the 
output in the case of $M^{({\rm pol)}}$ has an input-independent state 
of 
polarization, the intensity does depend on the state of polarization of 
the input. 
 
The matrix $M^{({\rm pin})}$  is not a Jones system, but it 
is a convex sum 
of such systems. To see this, note that a {\em perfect depolarizer} 
represented by the Mueller matrix
\begin{eqnarray}  
 M^{({\rm depol})} =\left[\begin{array}{cccc}
1 & 0 & 0 & 0\\
0 & 0 & 0 & 0\\
0   & 0 & 0 & 0\\
0   & 0 & 0 & 0
\end{array}\right],
\end{eqnarray}
is a convex sum of Jones systems; it can be realized,  
for instance, as equal mixture of systems with Jones matrices 
$\tau_a,\,a=0,\,1,\,2,\,3$. 
 That $M^{({\rm pin})}$ is a convex sum of 
Jones systems follows from 
$M^{({\rm pin})}=M^{({\rm pol})}M^{({\rm depol})}$.
 Alternatively, it is 
 readily seen that $H_{M^{({\rm pin})}}$ is an equal 
sum of two projections, and hence $M^{({\rm pin})}$  
is an equal mixture of the Jones systems 
\begin{eqnarray}
 \left[\begin{array}{cc}
1 & 0 \\
0 & 0 
\end{array}\right],~~~~~
 \left[\begin{array}{cc}
0 & 1 \\
0 & 0 
\end{array}\right].
\end{eqnarray}
While $M^{({\rm pin)}}$ is realized as convex sum of two Jones systems, 
$M^{({\rm depol)}}$ cannot be so realized with less than four Jones 
matrices. This follows from the fact that 
 $H_{M^{({\rm depol)}}}$  is of full rank whereas 
$H_{M^{({\rm pin)}}}$ is of rank two.   

The classification of canonical forms for $M$ matrices  
as given in (2.20) is complete in the following sense.

\noindent
{\bf Proposition}~3\,: Every $M$  matrix which maps the state space 
$\Omega^{({\rm pol})}$ into itself falls {\em uniquely} in one of 
the four families described in (2.20).

 That brings us to the {\em main thesis} of the present paper. A 
$4\times 4$  real 
matrix $M$ will have to map the state space 
$\Omega^{({\rm pol})}$ 
into itself in order that it qualifies to be the 
Mueller matrix of some physical system.  
This is certainly a necessary  condition.  
And, within the conventional Mueller-Stokes formalism, 
no conceivable further demand can be imposed on $M$. But the action of 
the 
{\em transversely homogeneous system represented by the numerical 
matrix} $M$ can be extended from plane waves to 
paraxial beams; naturally, $M$ 
will then affect only the polarization degree of freedom and {\em act 
as 
identity on the (transverse) spatial degrees of freedom}. If $M$ 
indeed  
represented a physical system, {\em even this extended action should 
map  
physical states into  physical states}. It turns out that this trivial 
looking extension is not all that trivial: there are $M$ matrices 
which appear physical at the level of the (restricted) state space 
 $\Omega^{({\rm pol})}$, but fail to be physical on the extended state 
space. Our task in the rest of the paper is to identify precisely those  
$M$ matrices whose action is physical even on the extended state space.  
 {\em Since only those $M$ matrices which pass this further hurdle 
can be 
called physical Mueller matrices, and pending determination of the 
precise demand this hurdle places on $M$, the $M$ matrices which map 
$\Omega^{({\rm pol})}$ into itself will be called  pre-Mueller 
matrices}. We may thus conclude this Section by saying that 
Eq.\,(2.20) gives a complete classification of pre-Mueller matrices    
 and their orbit structure    
under double-coseting by elements of $SO(3,1)$\,; the 
physical/unphysical 
divide of pre-Mueller matrices remains to  be accomplished. This 
divide  
will be presented in Section~4 
after some further preparation in Section~3.

\section{From Plane Waves to Beams: the BCP Matrix}
 
We will now go beyond plane waves and consider 
paraxial electromagnetic beams. 
The simplest 
(quasi-) monochromatic  beam field  has, in a transverse
plane $z=$ constant described by coordinates $(x,y)\equiv 
{\mbox{\boldmath$\rho$}},$ the
form 
$\mbox{\boldmath$E$}({\mbox{\boldmath$\rho$}})=(E_1\hat{\mbox{\boldmath$x$}}
+E_2\hat{\mbox{\boldmath$y$}})\,\psi({\mbox{\boldmath$\rho$}})$, 
where 
$E_1,\,E_2$ are  complex constants, and the scalar-valued function 
$\psi({\mbox{\boldmath$\rho$}})$ may be assumed to be square-integrable over
 the  transverse plane: $\psi({\mbox{\boldmath$\rho$}}) 
\in L^2({\mbox{\boldmath$R$}}^2)$. 
 It is clear that 
 the polarization part              
$(E_1\hat{\mbox{\boldmath$x$}}+E_2\hat{\mbox{\boldmath$y$}})$ and the 
spatial dependence or 
modulation
part $\psi({\mbox{\boldmath$\rho$}})$ of such a beam are well 
separated, 
allowing one to focus attention  
on one aspect
at a time. When one is interested in only the modulation aspect, 
the part
$(E_1\hat{\mbox{\boldmath$x$}}+E_2\hat{\mbox{\boldmath$y$}})$ may be 
suppressed, thus 
leading to `scalar optics'\,: this
is the domain of traditional Fourier optics\,\cite{Goodman}. 
[\,Fourier optics for  electromagnetic beams requires a 
 more delicate formalism\,\cite{Fourier85}.\,] On the 
other hand, if the 
spatial
part $\psi({\mbox{\boldmath$\rho$}})$ is suppressed we are led to the traditional
polarization optics (of plane waves) described in the previous Section.

Beams whose polarization and spatial modulation separate in the 
above manner will be called {\em elementary beams}. It is clear that 
elementary beams remain elementary under the action of transversely 
homogeneous anisotropic systems like waveplates and polarizers. That 
they remain elementary under the action of isotropic or polarization 
insensitive systems like free propagation and lenses is also 
clear.

Now suppose we superpose or add two such elementary beam fields
$(a\hat{\mbox{\boldmath$x$}}+b\hat{\mbox{\boldmath$y$}})\,
\psi({\mbox{\boldmath$\rho$}})$ and
$(c\hat{\mbox{\boldmath$x$}}+d\hat{\mbox{\boldmath$y$}})\,
\chi({\mbox{\boldmath$\rho$}})$. The 
result is not of the
elementary form $(e 
\hat{\mbox{\boldmath$x$}}+f\hat{\mbox{\boldmath$y$}})
\,\phi({\mbox{\boldmath$\rho$}})$,  for any
$e,~f,~\phi({\mbox{\boldmath$\rho$}}),$ unless either 
$(a,\,b)$ is proportional to $(c,\,d)$
 so that one gets committed to a {\em common} polarization, 
or $\psi({\mbox{\boldmath$\rho$}})$ and
$\chi({\mbox{\boldmath$\rho$}})$ are proportional so that one gets 
committed to a {\em fixed} 
spatial mode. In other words, {\em the set of elementary fields is not 
closed 
under superposition}.

{\em  Since one 
cannot possibly give up superposition principle in optics, one
 needs to go beyond  the set of  elementary fields and pay attention 
to the consequences of 
inseparability or entanglement of polarization and 
spatial variation (modulation)}.
 We are thus led to consider (in a  transverse plane) beam fields of 
the more general form
${\mbox{\boldmath$E$}}({\mbox{\boldmath$\rho$}})
=E_1({\mbox{\boldmath$\rho$}})\hat{\mbox{\boldmath$x$}}
+E_2({\mbox{\boldmath$\rho$}})\hat{\mbox{\boldmath$y$}}$.  
{\em This  form is
 obviously closed under superposition}. We may write 
${\mbox{\boldmath$E$}}({\mbox{\boldmath$\rho$}})$ as a 
{\em generalised Jones vector}
\begin{eqnarray}
{\mbox{\boldmath$E$}}({\mbox{\boldmath$\rho$}})
&=&E_1({\mbox{\boldmath$\rho$}})\hat{\mbox{\boldmath$x$}}
+E_2({\mbox{\boldmath$\rho$}})\hat{\mbox{\boldmath$y$}} 
\Leftrightarrow
 {\mbox{\boldmath$E$}}({\mbox{\boldmath$\rho$}})
=\left[\begin{array}{c}E_1({\mbox{\boldmath$\rho$}})\\
E_2({\mbox{\boldmath$\rho$}})
\end{array}\right],\nonumber\\
&&
~~E_1({\mbox{\boldmath$\rho$}}),\,\,
E_2({\mbox{\boldmath$\rho$}})
\in L^2({\mbox{\boldmath$R$}}^2)\,. 
\end{eqnarray}
The 
intensity at location 
${\mbox{\boldmath$\rho$}}$ corresponds to 
$|E_1({\mbox{\boldmath$\rho$}})|^2 + |E_2({\mbox{\boldmath$\rho$}})|^2$.
 This field is of the elementary or separable form iff
$E_1({\mbox{\boldmath$\rho$}})$ and $E_2({\mbox{\boldmath$\rho$}})$  
are linearly dependent (proportional to one another). Otherwise, 
polarization and spatial modulation are inseparably entangled. 

The point is that the set of possible beam fields in a transverse plane 
constitutes the {\em tensor product} space 
${\cal C}^2 \otimes L^2({\mbox{\boldmath$R$}}^2)$, 
 whereas the set of all elementary fields constitutes  
 just the {\em set product}  ${\cal C}^2 \times L^2({\mbox{\boldmath$R$}}^2)$  
of  ${\cal C}^2$  and $L^2({\mbox{\boldmath$R$}}^2)$.
  [Recall that the tensor product of two vector spaces is the 
closure of 
their set product under superposition.] Thus the set product 
${\cal C}^2 \times L^2({\mbox{\boldmath$R$}}^2)$    
forms a {\em measure zero subset} of the tensor product 
${\cal C}^2 \otimes L^2({\mbox{\boldmath$R$}}^2)$. 
 In other words, in a beam field represented by a {\em generic element} 
of ${\cal C}^2 \otimes L^2({\mbox{\boldmath$R$}}^2)$  polarization and 
 spatial modulation  
 should be expected to be  entangled\,:  {\em Entanglement is not an 
exception; 
it is the rule in} ${\cal C}^2 \otimes L^2({\mbox{\boldmath$R$}}^2)$, 
 the space of pure states appropriate for electromagnetic beams. 

If a beam described by generalised Jones vector 
$E({\mbox{\boldmath$\rho$}})$ with  $x,\,y$ components 
$E_1({\mbox{\boldmath$\rho$}}),\, 
E_2({\mbox{\boldmath$\rho$}})$ is 
passed through an $x$-polarizer, it is not only that the output will be   
$x$-polarized, it is certain to be in the spatial 
mode $E_1({\mbox{\boldmath$\rho$}})$ as well. 
Similar conclusion holds if the beam is passed through a $y$-polarizer. 
Thus a (transversely homogeneous) polarizer, whose action is 
${\mbox{\boldmath$\rho$}}$-independent, {\em not only chooses a 
polarization state, but acts 
as a spatial mode selector as well}.  This is true even if 
$E_1({\mbox{\boldmath$\rho$}})$ and $E_2({\mbox{\boldmath$\rho$}})$  
 are not spatially orthogonal modes. In a similar manner, 
{\em a spatial mode selector insensitive to polarization will end up 
acting 
also as a polarization discriminater}.  This is but one 
ramification of inseparability or entanglement between polarization and 
 spatial modulation.  
 
Now to handle fluctuating beams, we pass on to the 
beam-coherence-polarization (BCP) matrix 
$\Phi({\mbox{\boldmath$\rho$}};\,{\mbox{\boldmath$\rho$}}') \equiv
\langle {\mbox{\boldmath$E$}}({\mbox{\boldmath$\rho$}})
{\mbox{\boldmath$E$}}({\mbox{\boldmath$\rho$}}')^{\dagger}\rangle$,  
defined as the ensemble average of an
outer-product of (generalised) Jones 
vectors\,\cite{Gori98,Gorikids98}.  
As the name suggests, the 
BCP matrix describes both the
coherence and polarization properties of the beam under consideration. 
It is a
generalization of the numerical coherency matrix of plane waves 
considered in the previous Section, Eq.\,(2.2),  now to the case of 
beam fields. It can equally well be viewed as a generalization of the 
mutual 
coherence function of scalar statistical optics to include polarization.  
We are free to view the BCP matrix 
either as the $2\times 2$ matrix of two-point functions
\begin{eqnarray}
 \Phi({\mbox{\boldmath$\rho$}};{\mbox{\boldmath$\rho$}}')=
\left[\begin{array}{cc}
\langle E_1({\mbox{\boldmath$\rho$}})E_1({\mbox{\boldmath$\rho$}}')^*\rangle&
\langle E_1({\mbox{\boldmath$\rho$}})E_2({\mbox{\boldmath$\rho$}}')^*\rangle\\
\langle E_2({\mbox{\boldmath$\rho$}})E_1({\mbox{\boldmath$\rho$}}')^*\rangle &
\langle E_2({\mbox{\boldmath$\rho$}})E_2({\mbox{\boldmath$\rho$}}')^*\rangle
\end{array}
\right],
\end{eqnarray}
or as the associated column vector 
 $\tilde{\Phi}({\mbox{\boldmath$\rho$}};{\mbox{\boldmath$\rho$}}')$ 
of two-point functions\,:  
 $\tilde{\Phi}({\mbox{\boldmath$\rho$}};{\mbox{\boldmath$\rho$}}')
= \langle {\mbox{\boldmath$E$}}({\mbox{\boldmath$\rho$}})
\otimes {\mbox{\boldmath$E$}}({\mbox{\boldmath$\rho$}}')^{*}\rangle$.  
 For our present purpose, there is no need to make any finer distinction 
between the space-time and space-frequency descriptions. The 
two-point functions appearing in the BCP matrix may be viewed 
 either as equal-time coherence functions or correlation functions 
at a particular frequency. 

It is clear from the very definition (3.2) of  BCP matrix that 
this matrix kernel, viewed as an operator from 
${\cal C}^2\otimes L^2({\mbox{\boldmath$R$}}^2) 
\to {\cal C}^2\otimes L^2({\mbox{\boldmath$R$}}^2)$, 
  is hermitian and positive
semidefinite:
\begin{eqnarray}
\Phi_{jk}({\mbox{\boldmath$\rho$}};{\mbox{\boldmath$\rho$}}')
=\Phi_{kj}({\mbox{\boldmath$\rho$}}';{\mbox{\boldmath$\rho$}})^*,~~j,\,k=1,\,2;&&\nonumber\\
\int d^2{\mbox{\boldmath$\rho$}} \,d^2{\mbox{\boldmath$\rho$}}' 
{\mbox{\boldmath$E$}}({\mbox{\boldmath$\rho$}})^{\dagger}
{\Phi}({\mbox{\boldmath$\rho$}};{\mbox{\boldmath$\rho$}}')
{\mbox{\boldmath$E$}}({\mbox{\boldmath$\rho$}}')\ge 
0,&&\nonumber\\
{\rm i.e.},\; \sum_{jk} \int d^2{\mbox{\boldmath$\rho$}}\, 
d^2{\mbox{\boldmath$\rho$}}' 
E_j ({\mbox{\boldmath$\rho$}})^{*}
{\Phi}_{jk}({\mbox{\boldmath$\rho$}};{\mbox{\boldmath$\rho$}}')
E_k ({\mbox{\boldmath$\rho$}}')\ge 0,&&\nonumber\\ 
\forall {\mbox{\boldmath$E$}}({\mbox{\boldmath$\rho$}}) \in {\cal C}^2 
\otimes L^2({\mbox{\boldmath$R$}}^2).&& 
\end{eqnarray}
The positivity requirement thus demands that the expectation value of 
${\Phi}({\mbox{\boldmath$\rho$}};{\mbox{\boldmath$\rho$}}')$  be 
nonnegative for every Jones  vector 
${\mbox{\boldmath$E$}}({\mbox{\boldmath$\rho$}})$. 
 Hermiticity and positivity are the {\em defining
properties} of the BCP matrix: {\em every 
$2\times2$ matrix   
of two-point functions ${\Phi}_{jk}({\mbox{\boldmath$\rho$}};
{\mbox{\boldmath$\rho$}}')$ meeting just these
two conditions  is a valid BCP matrix of some beam of light}.

We can  use the BCP matrix to define,  in an obvious manner, the 
generalised  Stokes vector
$S({\mbox{\boldmath$\rho$}};{\mbox{\boldmath$\rho$}}')$\,\cite{Korotkova05}:
\begin{eqnarray}
{\Phi}({\mbox{\boldmath$\rho$}};{\mbox{\boldmath$\rho$}}')
&=&\frac{1}{2}\sum_{a=0}^3S_a({\mbox{\boldmath$\rho$}};
{\mbox{\boldmath$\rho$}}')\tau_a\nonumber\\ 
&&~~\Leftrightarrow~~ S_a({\mbox{\boldmath$\rho$}};
{\mbox{\boldmath$\rho$}}')={\rm 
tr}({\Phi}({\mbox{\boldmath$\rho$}};{\mbox{\boldmath$\rho$}}')\tau_a).~
\end{eqnarray}
That this is an {\em invertible} relation shows that 
${\Phi}({\mbox{\boldmath$\rho$}};{\mbox{\boldmath$\rho$}}')$
and $S({\mbox{\boldmath$\rho$}};{\mbox{\boldmath$\rho$}}')$ carry 
identical information: {\em action of an optical system on one 
defines a unique 
 equivalent action on the other}.  The 
hermiticity and
positivity requirement on the BCP matrix can be easily transcribed into
corresponding requirements on 
$S({\mbox{\boldmath$\rho$}};{\mbox{\boldmath$\rho$}}')$\,. Hermiticity 
reads 
\begin{eqnarray}
S_{a}({\mbox{\boldmath$\rho$}};{\mbox{\boldmath$\rho$}}')
=S_{a}({\mbox{\boldmath$\rho$}}';{\mbox{\boldmath$\rho$}})^*,
~~\,a=0,1,2,3;
\end{eqnarray}
whereas positivity reads 
\begin{eqnarray}
 \sum_{a=0}^3\,g_a\, \int d^2{\mbox{\boldmath$\rho$}}\, 
d^2{\mbox{\boldmath$\rho$}}' \,
{S}_{a}({\mbox{\boldmath$\rho$}};{\mbox{\boldmath$\rho$}}')
\widehat{S}_{a}({\mbox{\boldmath$\rho$}}';{\mbox{\boldmath$\rho$}})
 \ge 0, 
\end{eqnarray}
for every Stokes vector  
$\widehat{S}({\mbox{\boldmath$\rho$}};{\mbox{\boldmath$\rho$}}')$ 
arising from Jones vectors of the form 
${\mbox{\boldmath$E$}}({\mbox{\boldmath$\rho$}}) \in {\cal C}^2 
\otimes L^2({\mbox{\boldmath$R$}}^2)$. 
The signatures $g_a$ correspond to the `Lorentz metric': $g_0=1,\,\,g_a 
=-1$  for $a\ne1$.

\section{From pre-Mueller matrices to Mueller matrices: the role of 
entanglement}

We now have at our disposal all the tools necessary to determine 
if a given pre-Mueller matrix is a physical Mueller matrix or not. Let 
us  consider the transformation of the generalized Stokes vector  
$S({\mbox{\boldmath$\rho$}};{\mbox{\boldmath$\rho$}}')$ and the
 associated BCP matrix 
${\Phi}({\mbox{\boldmath$\rho$}};{\mbox{\boldmath$\rho$}}')$ 
under the action of a transversely homogeneous
 optical system described by  pre-Mueller matrix  
$M$.  We begin our analysis with pre-Mueller matrices of Type-I.

\subsection{Type-I pre-Mueller matrices}

We will first study  pre-Mueller matrices presented in the canonical 
form 
$M^{(1)} = {\rm diag}\,(d_0,\,d_1,\,d_2,\,d_3)$. 
Extension 
of the conclusions to pre-Mueller 
matrices not in the canonical form  will turn out to be quite 
straightforward.  
In  view of the system's homogeneity, 
the action of $M^{(1)}$ is {\em necessarily 
 independent} of ${\mbox{\boldmath$\rho$}},\,{\mbox{\boldmath$\rho$}}'$, 
and we have
\begin{eqnarray}
&&M^{(1)} = {\rm diag}\,(d_0,\,d_1,\,d_2,\,d_3):\nonumber\\ 
&&~~~\left[\begin{array}{c}
S_0({\mbox{\boldmath$\rho$}};{\mbox{\boldmath$\rho$}}')\\
S_1({\mbox{\boldmath$\rho$}};{\mbox{\boldmath$\rho$}}')\\
S_2({\mbox{\boldmath$\rho$}};{\mbox{\boldmath$\rho$}}')\\
S_3({\mbox{\boldmath$\rho$}};{\mbox{\boldmath$\rho$}}')
\end{array}\right] \to
\left[\begin{array}{c}
S_0'({\mbox{\boldmath$\rho$}};{\mbox{\boldmath$\rho$}}')\\
S_1'({\mbox{\boldmath$\rho$}};{\mbox{\boldmath$\rho$}}')\\
S_2'({\mbox{\boldmath$\rho$}};{\mbox{\boldmath$\rho$}}')\\
S_3'({\mbox{\boldmath$\rho$}};{\mbox{\boldmath$\rho$}}')
\end{array}\right]
=\left[\begin{array}{c}
d_0\,S_0({\mbox{\boldmath$\rho$}};{\mbox{\boldmath$\rho$}}')\\
d_1\,S_1({\mbox{\boldmath$\rho$}};{\mbox{\boldmath$\rho$}}')\\
d_2\,S_2({\mbox{\boldmath$\rho$}};{\mbox{\boldmath$\rho$}}')\\
d_3\,S_3({\mbox{\boldmath$\rho$}};{\mbox{\boldmath$\rho$}}')
\end{array}\right]\!.~\;\,\nonumber\\
\end{eqnarray}
The elements of the output BCP matrix 
$\Phi'({\mbox{\boldmath$\rho$}};{\mbox{\boldmath$\rho$}}')$ associated 
with the output Stokes vector 
$S'({\mbox{\boldmath$\rho$}};{\mbox{\boldmath$\rho$}}')$ 
resulting from the action of $M^{(1)}$ on 
$S({\mbox{\boldmath$\rho$}};{\mbox{\boldmath$\rho$}}')$, 
are easily computed using Eq.\,(3.4): 
\begin{eqnarray}
\Phi'_{11}({\mbox{\boldmath$\rho$}};{\mbox{\boldmath$\rho$}}') &=&
    [\,(d_0+d_1)\Phi_{11}({\mbox{\boldmath$\rho$}};{\mbox{\boldmath$\rho$}}')
    +(d_0-d_1)\Phi_{22}({\mbox{\boldmath$\rho$}};{\mbox{\boldmath$\rho$}}')
\,]/2,\nonumber\\
\Phi'_{22}({\mbox{\boldmath$\rho$}};{\mbox{\boldmath$\rho$}}') &=&
   [\,(d_0+d_1)\Phi_{22}({\mbox{\boldmath$\rho$}};{\mbox{\boldmath$\rho$}}')
    +(d_0-d_1)\Phi_{11}({\mbox{\boldmath$\rho$}};{\mbox{\boldmath$\rho$}}')
\,]/2,\nonumber\\
\Phi'_{12}({\mbox{\boldmath$\rho$}};{\mbox{\boldmath$\rho$}}') &=&
   [\,(d_2+d_3)\Phi_{12}({\mbox{\boldmath$\rho$}};{\mbox{\boldmath$\rho$}}')
    +(d_2-d_3)\Phi_{21}({\mbox{\boldmath$\rho$}};{\mbox{\boldmath$\rho$}}')
\,]/2,\nonumber\\
\Phi'_{21}({\mbox{\boldmath$\rho$}};{\mbox{\boldmath$\rho$}}') &=&
 [\,(d_2+d_3)\Phi_{21}({\mbox{\boldmath$\rho$}};{\mbox{\boldmath$\rho$}}')
    +(d_2-d_3)\Phi_{12}({\mbox{\boldmath$\rho$}};
{\mbox{\boldmath$\rho$}}')\,]/2. \nonumber\\
\end{eqnarray}

Clearly, a {\em necessary condition} for the pre-Mueller matrix 
$M^{(1)}={\rm 
diag}(d_0,\,d_1,\,d_2,\,d_3)$ to be  a physical Mueller 
matrix is that the output 
 $\Phi'({\mbox{\boldmath$\rho$}};{\mbox{\boldmath$\rho$}}')$ 
in Eq.\,(4.2) be a valid BCP matrix,  for  every
 valid input BCP matrix 
$\Phi({\mbox{\boldmath$\rho$}};{\mbox{\boldmath$\rho$}}')$. 
Hermiticity of
$\Phi'({\mbox{\boldmath$\rho$}};{\mbox{\boldmath$\rho$}}')$ 
is manifest in view of that of
$\Phi({\mbox{\boldmath$\rho$}};{\mbox{\boldmath$\rho$}}')$ 
and reality of the parameters $d_a$. Thus what remains 
to be checked is the positivity
of $\Phi'({\mbox{\boldmath$\rho$}};{\mbox{\boldmath$\rho$}}')$. 
While testing positivity of a  generic matrix kernel  
 could be a formidable task in general, 
it turns out that this test can be carried out
fairly easily in the present case.

Let us take as the input a special 
pure state BCP matrix  
$\Phi^{(0)}({\mbox{\boldmath$\rho$}};{\mbox{\boldmath$\rho$}}') =
{\mbox{\boldmath$E$}}({\mbox{\boldmath$\rho$}})
{\mbox{\boldmath$E$}}({\mbox{\boldmath$\rho$}}')^{\,\dagger}$,  
corresponding to the 
generalised Jones vector
${\mbox{\boldmath$E$}}({\mbox{\boldmath$\rho$}})$ 
which is an equal superposition of an $x$-polarized mode and
a $y$-polarized mode, {\em the two modes being  spatially  
orthogonal}\,:
\begin{eqnarray}
\Phi^{(0)}({\mbox{\boldmath$\rho$}};{\mbox{\boldmath$\rho$}}') &=&
{\mbox{\boldmath$E$}}({\mbox{\boldmath$\rho$}})
{\mbox{\boldmath$E$}}({\mbox{\boldmath$\rho$}}')^{\,\dagger}\,,\nonumber\\ 
{\mbox{\boldmath$E$}}({\mbox{\boldmath$\rho$}})
=\left[\begin{array}{c}\psi_1({\mbox{\boldmath$\rho$}})\\
\psi_2({\mbox{\boldmath$\rho$}})
  \end{array}
  \right],&&\!\!\int\psi_j({\mbox{\boldmath$\rho$}})
\psi_k({\mbox{\boldmath$\rho$}})^*d^2{\mbox{\boldmath$\rho$}} = 
\delta_{jk}.~\;\;
\end{eqnarray}
[\,$\psi_1({\mbox{\boldmath$\rho$}})$ and 
$\psi_2({\mbox{\boldmath$\rho$}})$ 
could, for instance, be two distinct 
Hermite-Gaussian modes.\,] 
This means that the entries of the input  BCP matrix 
$\Phi^{(0)}({\mbox{\boldmath$\rho$}};{\mbox{\boldmath$\rho$}}')$ 
have the deterministic form 
$\Phi^{(0)}_{jk}({\mbox{\boldmath$\rho$}};{\mbox{\boldmath$\rho$}}')
=\psi_j({\mbox{\boldmath$\rho$}})\psi_k({\mbox{\boldmath$\rho$}}')^*$. 
 A consequence of this simple (product) form is that {\em the four entries 
of the BCP matrix 
$\Phi^{(0)}({\mbox{\boldmath$\rho$}};{\mbox{\boldmath$\rho$}}')$ 
form an orthonormal set of  (two-point) functions}\,: 
\begin{eqnarray}
\int d^2 {\mbox{\boldmath$\rho$}}\,d^2 {\mbox{\boldmath$\rho$}}'
\Phi^{(0)}_{ij}({\mbox{\boldmath$\rho$}};{\mbox{\boldmath$\rho$}}')
\Phi^{(0)}_{kl}({\mbox{\boldmath$\rho$}};{\mbox{\boldmath$\rho$}}')^{*} 
=\delta_{ik}\delta_{jl}.
\end{eqnarray}
This fact will prove to be of much value in our analysis below. 

To test  positivity of the output BCP matrix 
$\Phi'({\mbox{\boldmath$\rho$}};{\mbox{\boldmath$\rho$}}')$,  
given in Eq.\,(4.2) and resulting from input 
$\Phi^{(0)}({\mbox{\boldmath$\rho$}};{\mbox{\boldmath$\rho$}}')
={\mbox{\boldmath$E$}}({\mbox{\boldmath$\rho$}})
{\mbox{\boldmath$E$}}({\mbox{\boldmath$\rho$}}')^{\dagger}$, let us
define four (generalized) Jones vectors:
\begin{eqnarray}
 {\mbox{\boldmath$E$}}^{(\pm)}({\mbox{\boldmath$\rho$}})=
\left[\begin{array}{c}\psi_1({\mbox{\boldmath$\rho$}})\\\pm\, 
\psi_2({\mbox{\boldmath$\rho$}})
 
\end{array}\right],~\,{\mbox{\boldmath$F$}}^{(\pm)}({\mbox{\boldmath$\rho$}})=
\left[\begin{array}{c}\psi_2({\mbox{\boldmath$\rho$}})\\
\pm\, \psi_1({\mbox{\boldmath$\rho$}})
\end{array}\right].~
\end{eqnarray}
  [\,The input Jones vector 
${\mbox{\boldmath$E$}}({\mbox{\boldmath$\rho$}})$ happens to coincide with  
 ${\mbox{\boldmath$E$}}^{(+)}({\mbox{\boldmath$\rho$}})$\,].  
 Expectation values of 
$\Phi'({\mbox{\boldmath$\rho$}};{\mbox{\boldmath$\rho$}}')$ for  the  four Jones 
vectors  ${\mbox{\boldmath$E$}}^{(\pm )}({\mbox{\boldmath$\rho$}})
,\; {\mbox{\boldmath$F$}}^{(\pm )}({\mbox{\boldmath$\rho$}})$ 
are easily computed using Eqs.\,(4.2),\,(4.3),\,(4.4) and (4.5) in 
Eq.\,(3.3). 
 These expectation values are $(d_0 + d_1) \pm (d_2 + d_3)$ for  
 ${\mbox{\boldmath$E$}}^{(\pm )}({\mbox{\boldmath$\rho$}})$ 
and $(d_0 - d_1) \pm (d_2 - d_3)$ for  
${\mbox{\boldmath$F$}}^{(\pm )}({\mbox{\boldmath$\rho$}})$.  

Now positivity of 
$\Phi'({\mbox{\boldmath$\rho$}};{\mbox{\boldmath$\rho$}}')$ 
requires, as a necessary condition,  that these four expectation 
values be nonnegative, and this demand places on the parameters $d_a$ 
the constraints  
\begin{eqnarray}
-d_1-d_2-d_3 &\le& d_0,\nonumber\\
-d_1+d_2+d_3  &\le& d_0;\nonumber\\
 d_1+d_2-d_3 &\le& d_0,\nonumber\\
d_1-d_2+d_3  &\le& d_0.
\end{eqnarray}
Violation of any one of these four conditions will render the output  
 $\Phi'({\mbox{\boldmath$\rho$}};{\mbox{\boldmath$\rho$}}')$ unphysical 
as BCP matrix. Since the input BCP matrix 
$\Phi^{(0)}({\mbox{\boldmath$\rho$}};{\mbox{\boldmath$\rho$}}')$ 
  is  obviously physical, this will in turn render 
$M^{(1)}$ unphysical as Mueller matrix: 
Eq.\,(4.6) is thus a {\em set of necessary conditions} for  
 the pre-Mueller matrix $M^{(1)}$ to be a Mueller matrix. 

Suppose these four inequalities are met. Can we then conclude that 
 the pre-Mueller matrix $M^{(1)}$ is a physically acceptable 
Mueller 
matrix? To answer this question in the affirmative we write in detail 
the associated 
hermitian 
matrix $H_{M^{(1)}} = \frac{1}{2}\sum_a\,d_a\,\tau_a\otimes\tau_a^{*}
=\sum_a d_aU_{aa}$\,:
\begin{eqnarray}
H_M =\frac{1}{2} 
\left[\begin{array}{cccc} 
d_0+ d_1\, &
0&
0 &
d_2 + d_3\, \\
0 &
\,d_0  - d_1\, &
\,d_2 - d_3\,  &
0 \\
0 &
\,d_2 - d_3\,  &
\,d_0   - d_1\, &
0 \\
\,d_2 + d_3  &
0 &
0 &
\,d_0   + d_1 
\end{array}\right]\!. \;\;
\end{eqnarray}
Validity of the four inequalities in Eq.\,(4.6)
implies, fortunately, that this matrix is positive semidefinite.  
 This in turn implies that the given diagonal   
system $M^{(1)}$ is a convex sum of Jones systems, and therefore 
takes every BCP matrix into a BCP matrix, showing that 
Eq.\,(4.6) is  {\em sufficient condition} for 
$M^{(1)}$ to be a Mueller matrix. We have thus proved 

\noindent
{\bf Proposition}~4\,: The  pre-Mueller matrix    
 $M^{(1)} = {\rm diag}\,(d_0,\,d_1,\,d_2,\,d_3)$ is a Mueller matrix 
iff the associated hermitian matrix $H_{M^{(1)}} \ge 0$. That is,   
iff $M^{(1)}$ can be realized as a convex sum of Jones systems or,
equivalently, iff the entries of $M^{(1)}$ 
respect the inequalities in (4.6). 

Having settled the diagonal case, we  now go beyond and consider 
 non-diagonal Type-I pre-Mueller matrices. 
 As noted in (2.20), these are 
necessarily of the  general form 
$M= L_{\ell}M^{(1)}L_r$, where 
$L_{\ell},\,L_r \in SO(3,1)$ and $M^{(1)}$ is diagonal. We have already 
noted that $L_{\ell},\,L_r$ are physical Mueller matrices: indeed, they 
correspond to deterministic systems with  respective 
Jones matrices $J_{\ell},\,J_r\in SL(2,C)$.    
 Thus if $M^{(1)}$ is a Mueller matrix, then  it has a 
convex sum realization $\{\,p_k,\,J^{(k)}\,\}$. This implies  
that  $M= L_{\ell}M^{(1)}L_r$ has the convex sum realization 
 $\{\,p_k,\,J_{\ell}J^{(k)}J_r\,\}$, and hence is a valid Mueller 
matrix. The converse  follows by  virtue of the 
{\em invertibility} 
of     $J_{\ell},\,J_r$, and we have     
  
\noindent
{\bf Proposition} 5\,: A Type-I pre-Mueller matrix, which is 
necessarily of the form $M= 
L_{\ell}M^{(1)}L_r$   with 
$L_{\ell},\,L_r \in SO(3,1)$ and $M^{(1)}$ diagonal, 
is a physical Mueller matrix iff $M^{(1)}$ is.

\subsection{Type-II pre-Mueller matrices}

Having fully classified Type-I pre-Mueller matrices into 
Mueller  and non-Mueller matrices, we now turn our attention to 
Type-II pre-Mueller matrices. The analysis turns out to be quite 
parallel 
to the 
one in the previous sub-Section. 

Recall from Section~2 that a Type-II pre-Mueller matrix in its 
canonical 
form 
$M^{(2)}$ has only one nonvanishing off-diagonal element whose value is 
fixed by the diagonals, namely $M_{01} = d_0-d_1$, where 
$d_0,\,d_1,\,d_2,\,d_3$ are the diagonals.   
 The action of  $M^{(2)}$ on 
$S({\mbox{\boldmath$\rho$}};{\mbox{\boldmath$\rho$}}')$  
and
$\Phi({\mbox{\boldmath$\rho$}};{\mbox{\boldmath$\rho$}}')$ 
can be computed as before. The (generalised) Stokes vector has this 
simple transformation law: 

\begin{eqnarray} 
M^{(2)}:&&S({\mbox{\boldmath$\rho$}};{\mbox{\boldmath$\rho$}}')\,\to\, 
S'({\mbox{\boldmath$\rho$}};{\mbox{\boldmath$\rho$}}')
=
\left[\begin{array}{c} 
S_0'({\mbox{\boldmath$\rho$}};{\mbox{\boldmath$\rho$}}')\\ 
S_1'({\mbox{\boldmath$\rho$}};{\mbox{\boldmath$\rho$}}')\\ 
S_2'({\mbox{\boldmath$\rho$}};{\mbox{\boldmath$\rho$}}')\\ 
S_3'({\mbox{\boldmath$\rho$}};{\mbox{\boldmath$\rho$}}') 
\end{array}\right] \nonumber\\
&=&\left[\begin{array}{c} 
d_0\,S_0({\mbox{\boldmath$\rho$}};{\mbox{\boldmath$\rho$}}')
+(d_0 - d_1)\,S_1({\mbox{\boldmath$\rho$}};{\mbox{\boldmath$\rho$}}')\\ 
d_1\,S_1({\mbox{\boldmath$\rho$}};{\mbox{\boldmath$\rho$}}')\\ 
d_2\,S_2({\mbox{\boldmath$\rho$}};{\mbox{\boldmath$\rho$}}')\\ 
d_3\,S_3({\mbox{\boldmath$\rho$}};{\mbox{\boldmath$\rho$}}')
\end{array}\right]\!.\;\;\, 
\end{eqnarray} 

The elements of the   output BCP matrix 
$\Phi'({\mbox{\boldmath$\rho$}};{\mbox{\boldmath$\rho$}}')$ 
 associated with 
$S'({\mbox{\boldmath$\rho$}};{\mbox{\boldmath$\rho$}}')$ 
 and computed from (3.4) are 
\begin{eqnarray} 
\Phi'_{11}({\mbox{\boldmath$\rho$}};{\mbox{\boldmath$\rho$}}') &=&
 d_0\,\Phi_{11}({\mbox{\boldmath$\rho$}};{\mbox{\boldmath$\rho$}}')\,,\nonumber\\
\Phi'_{22}({\mbox{\boldmath$\rho$}};{\mbox{\boldmath$\rho$}}') &=&
   d_1\,\Phi_{22}({\mbox{\boldmath$\rho$}};{\mbox{\boldmath$\rho$}}')
 +(d_0-d_1)\Phi_{11}({\mbox{\boldmath$\rho$}};{\mbox{\boldmath$\rho$}}')
\,,\nonumber\\
\Phi'_{12}({\mbox{\boldmath$\rho$}};{\mbox{\boldmath$\rho$}}') &=&
   [\,(d_2+d_3)\Phi_{12}({\mbox{\boldmath$\rho$}};{\mbox{\boldmath$\rho$}}')
    +(d_2-d_3)\Phi_{21}({\mbox{\boldmath$\rho$}};{\mbox{\boldmath$\rho$}}')
\,]/2,\nonumber\\
\Phi'_{21}({\mbox{\boldmath$\rho$}};{\mbox{\boldmath$\rho$}}') &=&
 [\,(d_2+d_3)\Phi_{21}({\mbox{\boldmath$\rho$}};{\mbox{\boldmath$\rho$}}')
    +(d_2-d_3)\Phi_{12}({\mbox{\boldmath$\rho$}};
{\mbox{\boldmath$\rho$}}')\,]/2. \nonumber\\
\end{eqnarray}
As in the case of $M^{(1)}$,  the canonical form 
pre-Mueller matrix  $M^{(2)}$ does not couple the pair   
$\Phi_{11}({\mbox{\boldmath$\rho$}};{\mbox{\boldmath$\rho$}}')$, 
$\Phi_{22}({\mbox{\boldmath$\rho$}};{\mbox{\boldmath$\rho$}}')$ with 
$\Phi_{12}({\mbox{\boldmath$\rho$}};{\mbox{\boldmath$\rho$}}')$, 
$\Phi_{21}({\mbox{\boldmath$\rho$}};{\mbox{\boldmath$\rho$}}')$.

Again, a necessary condition for the pre-Mueller matrix $M^{(2)}$ to be 
a physically acceptable Mueller matrix  
is that the output   
$\Phi'({\mbox{\boldmath$\rho$}};{\mbox{\boldmath$\rho$}}')$ 
in Eq.\,(4.9) be a valid BCP matrix for  every
 valid input BCP matrix 
$\Phi({\mbox{\boldmath$\rho$}};{\mbox{\boldmath$\rho$}}')$. 
 As in the case of $M^{(1)}$, let us  take as  input the   
pure state BCP matrix  
$\Phi^{(0)}({\mbox{\boldmath$\rho$}};{\mbox{\boldmath$\rho$}}') =
{\mbox{\boldmath$E$}}({\mbox{\boldmath$\rho$}})
{\mbox{\boldmath$E$}}({\mbox{\boldmath$\rho$}}')^{\,\dagger}$, with 
${\mbox{\boldmath$E$}}({\mbox{\boldmath$\rho$}})$ as described in 
Eq.\,(4.3). 
 To test  positivity of the output BCP matrix 
$\Phi'({\mbox{\boldmath$\rho$}};{\mbox{\boldmath$\rho$}}')$,  
given in Eq.\,(4.9) and resulting from input 
$\Phi^{(0)}({\mbox{\boldmath$\rho$}};{\mbox{\boldmath$\rho$}}')
={\mbox{\boldmath$E$}}({\mbox{\boldmath$\rho$}})
{\mbox{\boldmath$E$}}({\mbox{\boldmath$\rho$}}')^{\dagger}$, we use 
in place of  ${\mbox{\boldmath$E$}}^{(\pm)}({\mbox{\boldmath$\rho$}})$ 
and ${\mbox{\boldmath$F$}}^{(\pm)}({\mbox{\boldmath$\rho$}})$ slightly 
modified (generalized) Jones vectors
 ${\mbox{\boldmath$E$}}^{(\theta)}({\mbox{\boldmath$\rho$}})$ and 
 ${\mbox{\boldmath$F$}}^{(\theta)}({\mbox{\boldmath$\rho$}})$\,:
\begin{eqnarray}
 {\mbox{\boldmath$E$}}^{(\theta)}({\mbox{\boldmath$\rho$}})=
\left[\begin{array}{c}\cos \theta\,\psi_1({\mbox{\boldmath$\rho$}})\\
\sin \theta\, \psi_2({\mbox{\boldmath$\rho$}})
 \end{array}\right]\!,
\;{\mbox{\boldmath$F$}}^{(\theta)}({\mbox{\boldmath$\rho$}})=
\left[\begin{array}{c}\cos \theta\,\psi_2({\mbox{\boldmath$\rho$}})\\
\sin \theta\, \psi_1({\mbox{\boldmath$\rho$}})
\end{array}\right]\!.~\nonumber\\
\end{eqnarray}
 Expectation values of the output BCP matrix  
$\Phi'({\mbox{\boldmath$\rho$}};{\mbox{\boldmath$\rho$}}')$ 
for  these two families of  Jones 
vectors  can be computed as before.
 These expectation values are $d_0\cos ^2 \theta + d_1\sin ^2 \theta 
+ (d_2 + d_3)\cos \theta \sin \theta$ 
for  
 ${\mbox{\boldmath$E$}}^{\theta}({\mbox{\boldmath$\rho$}})$ 
and $(d_0 - d_1)\sin ^2 \theta + (d_2 - d_3)\cos \theta \sin \theta$ for  
${\mbox{\boldmath$F$}}^{\theta}({\mbox{\boldmath$\rho$}})$.

 Now positivity of 
$\Phi'({\mbox{\boldmath$\rho$}};{\mbox{\boldmath$\rho$}}')$ 
requires, as a necessary condition,  that these expectation 
values be nonnegative for all  $0\le \theta<\pi$\,, 
and this requirement is seen to be equivalent to the pair of conditions  
$d_0d_1\ge (d_2+d_3)^2/4$,\, $d_2-d_3=0$; these arise respectively 
from the   ${\mbox{\boldmath$E$}}^{(\theta)}$ and 
${\mbox{\boldmath$F$}}^{(\theta)}$ families. 
We may rewrite these as   
\begin{eqnarray}
 d_3 = d_2,\;\; (d_2)^2 \le d_0d_1\,. 
\end{eqnarray}
This is a pair of {\em necessary conditions} for 
 the pre-Mueller matrix $M^{(2)}$ 
to be a Mueller matrix.
 The condition $(d_2)^2 \le d_0d_1$ is already part of 
the definition of $M^{(2)}$, and thus 
$ d_3 = d_2$ is the  new requirement arising from consideration of 
the action of $M^{(2)}$ on BCP matrices, i.e., from consideration of 
entanglement.  

Our next task is to show that these conditions are sufficient as well.
 To this end  we proceed as in the case of $M^{(1)}$ and compute the  
hermitian 
matrix $H_{M^{(2)}}$ associated with  $M^{(2)}$\,:
\begin{eqnarray}
H_{M^{(2)}} =
\left[\!\begin{array}{cccc} 
d_0&
0&
0 &
\frac{1}{2}(d_2 + d_3)\\
0 &
0 &
\frac{1}{2}(d_2 - d_3)  &
0 \\
0 &
\frac{1}{2}(d_2 - d_3)  &
d_0   - d_1 &
0 \\
\frac{1}{2}(d_2 + d_3)  &
0 &
0 &
  d_1 
\end{array}\!\right]. \nonumber\\
\end{eqnarray}
The inequalities in Eq.\,(4.11) are precisely the conditions under 
which 
$H_{M^{(2)}}$ is positive.
 This in turn implies that   $M^{(2)}$ satisfying (4.11) 
 is a convex sum of Jones systems, and  
  hence is a Mueller matrix. We have thus proved 

\noindent
{\bf Proposition} 6\,: The  pre-Mueller matrix    
 $M^{(2)}$ is a Mueller matrix 
iff the associated hermitian matrix $H_{M^{(2)}} \ge 0$. That is,   
iff $M^{(2)}$ can be realized as a convex sum of Jones systems or,
equivalently, iff the entries of $M^{(2)}$ 
respect the inequalities in (4.11). 

We can now proceed to consider Type-II pre-Mueller matrices 
which are not of the canonical form $M^{(2)}$. We know 
from Section~2 that any such matrix  
 has the  form 
$M= L_{\ell}M^{(2)}L_r$, where 
$L_{\ell},\,L_r \in SO(3,1)$. By considerations similar to the ones 
 leading to Proposition~6 in the Type-I case,  we arrive at 
  
\noindent
{\bf Proposition} 7\,: A type-II pre-Mueller matrix,  
which is necessarily of the form $M= L_{\ell}M^{(2)}L_r$  with 
$L_{\ell},\,L_r \in SO(3,1)$,  is a Mueller matrix iff $M^{(2)}$ is. 

Having completed classification of the pre-Mueller matrices 
in the Type-I and Type-II families into physical and non-physical ones, 
we are now left with two minor families to handle. As noted following 
(2.20), $M^{({\rm pol})}$ is a Jones system. Let $J$ be the Jones 
matrix representing this system (polariser). In view of the two-to-one 
homomorphism between $SL(2,C)$ and $SO(3,1)$ alluded to earlier, 
$L_\ell,\,L_r\in SO(3,1)$ define respective Jones matrices 
$J_\ell,\,J_r$ of unit determinant; these Jones matrices are 
 {\em unique except for  multiplicative factor}  
$\pm 1$ and, as is well known, this 
signature ambiguity is of nontrivial origin. Thus $M=L_\ell M^{({\rm 
pol })}L_r$  is a Jones system with Jones matrix $\pm J_\ell JJ_r$, and 
hence is physical. Similar argument will show that the last family, 
namely the PinMap family, too has no non-physical $M$ matrix.  
 For completeness, we state the situation in respect of these two 
minor families as the following \\

\noindent
{\bf Proposition}~8\,: Pre-Mueller matrices belonging to the 
Polarizer and Pin\,Map families are respectively Jones systems and 
convex sums of Jones systems. Their associated $H$ matrices are 
positive semidefinite, and  all pre-Mueller matrices in these 
two families are physical Mueller matrices.

\subsection{Complete Characterization of Mueller Matrices}

 In the last two sub-Sections we carried out a complete 
classification of  
pre-Mueller matrices 
into physical and non-physical ones. 
Double-coseting under the 
$SO(3,1)$ group has played such an important role in this process 
that we capture this role as a separate result. 

\noindent
{\bf Proposition}~9\,: Given two  $4\times 4$ real matrices $M$ and  
$M'$ which are in the same double-coset orbit under $SO(3,1)$, i.e., 
$M'= L_\ell M L_r$ for  some $L_\ell,\,L_r\in SO(3,1)$,  
$M'$ is a convex sum of Jones systems iff $M$ is. And 
$H_{M'} \ge 0$  iff $H_{M} \ge 0$. In other words, $M'$ is 
a Mueller matrix iff $M$ is. 

\noindent
{\em Proof}\,: Suppose $M$ has the convex sum realization 
$\{p_k,\,J^{(k)}\}$, i.e., $M= \sum_k p_k\,M(J^{(k)})$. Then, clearly,  
 $M'$ has the convex sum realization 
$\{p_k,\,J_\ell J^{(k)}J_r\}$. 
 Conversely, if $M'$ has the convex sum realization 
$\{p'_k,\,J'^{\,(k)}\}$, then   
$M$ has the convex sum realization 
$\{p'_k,\,(J_\ell)^{-1} J'^{\,(k)}(J_r)^{-1}\}$. 

Now suppose $H_M\ge0$. This means  
$H_M = \sum_k p_k\,\tilde{J}^{(k)\,}\tilde{J}^{(k)\,\dagger},\;p_k > 0$.
 This immediately implies 
$H_{M'} = \sum_k p_k\,
\widetilde{(J_\ell J^{(k)}J_r)}
\widetilde{(J_\ell J^{(k)}J_r)}^{\,\dagger}$, which proves its 
positivity. Here $\widetilde{(J_\ell J^{(k)}J_r)}$, as usual, denotes 
the column vector associated with the $2\times 2$ 
matrix ${J_\ell J^{(k)}J_r}$. 
The converse follows from the invertibility of $J_\ell,\,J_r$, 
completing proof of the Proposition. 

 With this  proof of the 
 principal conclusion of this paper is complete.  Our main theorem may 
thus be  stated as  follows.  

\noindent
{\bf Theorem}\,: A $4\times 4$ real matrix $M$ is a Mueller matrix 
iff the associated hermitian matrix $H_M \ge0$. Every physically 
acceptable Mueller matrix is a convex sum of Mueller-Jones matrices. 

\subsection{The role of entanglement\,: An illustrative example}

We present a simple example to illustrate the kind or restrictions 
on $M$ matrices brought in by consideration of entanglement. Let 
us restrict attention to  $M$ matrices of the special simple 
 three-parameter form 
\begin{eqnarray}
M =
\left[\begin{array}{cccc} 
1&0&0&0\\
0&d_1&0&0\\
0&0&d_3&0\\
0&0&0&d_3
\end{array}\right]. 
\end{eqnarray}
We are obviously in the Type-I situation, but 
we are not considering here the $SO(3,1)$ 
orbit under double-coseting.

It is clear that $M$ will map Stokes vectors into Stokes vectors 
if and only if $M$ satisfies the following three conditions: 
\begin{eqnarray}
M:\; \Omega ^{({\rm pol})} &\to& \Omega ^{({\rm pol})} 
\Leftrightarrow 
\nonumber\\
&&- 1\, \le\, d_k \,\le \,1,\;\;\; k=1,\,2,\,3\,. 
\end{eqnarray}
{\em In the absence of considerations of entanglement these would have 
been 
the  only conditions  $M$ will need to satisfy}. Thus the allowed 
region in the Euclidean space 
$\mbox{\boldmath$R$}^3$ 
spanned by the parameters $(d_1,\,d_2,\,d_3)$ 
would have been the cube with vertices at $(\pm1,\,\pm1,\,\pm1)$.  

 Now each one of the four conditions in (4.6) with $d_0=1$, arising out 
of consideration of entanglement,   forbids the (open) half-space on one 
side 
of a plane. {\em Thus the allowed region is the intersection of 
the 
four allowed half-spaces}. This region is clearly the tetrahedron  
with vertices at $(1,\,1,\,1),\,(1,\,-1,\,-1),\,(-1,\,1,\,-1)$ and 
$(-1,\,-1,\,1)$.   

Thus, it is not the entire cubical region (4.14) permitted by 
conventional 
wisdom, but only 
the solid tetrahedron, with one third volume of the cube, that 
 stands the closer scrutiny presented by consideration of entanglement. 
 This is illustrated in Fig.1. 
The region outside the tetrahedron is unphysical and does not correspond 
to Mueller matrices: points in the cube but exterior to the tetrahedron 
correspond to pre-Mueller matrices which are not physical Mueller 
matrices; those outside the cube correspond to $M$ matrices which are 
not even pre-Mueller matrices.
 
\section{Concluding Remarks}
 
We conclude with some further observations. In 
the mathematics 
literature and in the literature of quantum information 
theory,  what we have called pre-Mueller matrices 
go by the name {\em positive maps}, and  the subset of pre-Mueller  
matrices which are physically acceptable 
in the sense of our main theorem corresponds to what are called 
{\em completely positive maps}. 
But we have endevoured here  to arrive at a physical 
characterization 
of Mueller matrices entirely within the  framework of  BCP matrices  
familiar   to the classical optics community,  without 
resorting to the mathematical theory of these maps.

Secondly, a pre-Mueller matrix which fails our main theorem {\em will 
not}  produce any unphysical effect acting on 
 the coherency matrix of plane waves or on the BCP matrix 
of elementary (or polarization-modulation separable) beams.
It follows that BCP matrices which are convex sums of elementary beams 
will not be able to witness the failure of a pre-Mueller matrix $M$ 
whose associated $H_M$ is not positive semidefinite. 
Only  BCP matrices corresponding to entangled generalized Jones vectors 
can expose the unphysical  
nature of a pre-Mueller matrix which violates  our main theorem.  In 
other words, {\em pre-Mueller matrices cannot be further divided into 
 physical and unphysical subsets without consideration of 
entanglement}.  
 
Finally, ever since it was proved that every Jones system corresponds to 
 a Mueller matrix whose associated $H$ matrix is a  
projection\,\cite{Simon82}, it 
has been clear that ensembles of Jones systems necessarily correspond to 
positive semidefinite $H$ matrices, and conversely. 
 It has thus been occationally suggested by various authors, beginning 
with \cite{Kim-Mandel-Wolf}, that 
considerations of Mueller matrices might be restricted 
 to only such  ensembles. 
But it has remained only a suggestion, and one without 
any physical basis, and hence could not set aside as unphysical an  
experimentally measured Mueller matrix whose associated $H$ matrix 
has a negative eigenvalue, {\em particularly when the reported Mueller 
system was not realized by the experimenter specifically as an ensemble 
of Jones systems}. For instance the symmetric Mueller matrix of van 
Zyl\,\cite{Zyl87}  was analysed in Ref.\,\cite{Sridhar94}, and was 
found to 
be a matrix of Type-I, with canonical-form parameters 
$d_0=0.9735,\;d_1=0.9112,\;d_2=0.4640, \;d_3= -0.3838$. This clearly 
violates [the second constraint in] Eq.\,(4.6) by a substantial 
 extent.  Equivalently, the eigenvalues of 
$H_M$ are $1.0906,\;0.8393,\;0.4526,\; -0.3825$. 
 
That $H_M$ in this case is not positive, 
and hence the van Zyl system is not a convex sum 
of Jones systems was always known.   
However, one did not have hitherto  a 
physical basis on which this $M$ could be judged as unphysical. 
Non-quantum entanglement has now given us such a physical basis.\\

\noindent
{\bf Acknowledgements}: This work was presented by R. Simon at the Koli
Workshop on {\em Partial Electromagnetic Coherence and 3D Polarization}
May 24--27, 2009, Koli, Finland. He would like to thank Ari Friberg, Jari
Turunen, and Jani Tervo for making his participation possible.\\

R. Simon's e-mail address is simon@imsc.res.in

\end{document}